\documentclass[prl,aps,twocolumn,showpacs,10pt]{revtex4-1}
\usepackage{graphicx}
\usepackage{amsmath}
\usepackage[table]{xcolor}

\begin{document}

\title{Space Charge Transfer in Hybrid Inorganic/Organic Systems}

\author{Yong \surname{Xu}$^1$}
\email{yongxu@fhi-berlin.mpg.de}
\author{Oliver T.\ \surname{Hofmann}$^1$}
\author{Raphael \surname{Schlesinger}$^2$}
\author{Stefanie \surname{Winkler}$^3$}
\author{Johannes \surname{Frisch}$^2$}
\author{Jens \surname{Niederhausen}$^2$}
\author{Antje \surname{Vollmer}$^3$}
\author{Sylke \surname{Blumstengel}$^2$}
\author{Fritz \surname{Henneberger}$^2$}
\author{Norbert \surname{Koch}$^{2,3}$}
\author{Patrick \surname{Rinke}$^1$}
\author{Matthias \surname{Scheffler}$^1$}

\affiliation{$^1$Fritz-Haber-Institut der Max-Planck-Gesellschaft, 14195 Berlin, Germany \\
$^2$Humboldt-Universit\"at zu Berlin, Institut f\"ur Physik, 12489 Berlin, Germany \\
$^3$Helmholtz-Zentrum Berlin f\"ur Materialien und Energie GmbH-BESSY II, 12489 Berlin, Germany}


\begin{abstract}
We discuss density functional theory calculations of hybrid inorganic/organic systems (HIOS) that explicitly include the global effects of doping (i.e. position of the Fermi level) and the formation of a space-charge layer. For the example of tetrafluoro-tetracyanoquinodimethane (F4TCNQ) on the ZnO(000$\bar{1}$) surface we show that the adsorption energy and electron transfer depend strongly on the ZnO doping. The associated work function changes are large, for which the formation of space-charge layers is the main driving force. The prominent doping effects are expected to be quite general for charge-transfer interfaces in HIOS and important for device design.
\end{abstract}



\maketitle

Hybrid inorganic/organic systems (HIOS) have already been applied in (opto)electronics, including solar cells~\cite{Law-05}, laser diodes~\cite{Yang-08}, light emitting diodes~\cite{Sessolo-11} or sensors~\cite{Levell-10}. Recently HIOS have attracted enormous research interest owing to their promise to synergetically combine the best features of two worlds. This could be, for example, the high charge carrier mobility and efficient charge injection of inorganic semiconductors, and the strong light-matter coupling and large chemical compound space of organic semiconductors.

\begin{figure}[tbhp]
\includegraphics[width=0.8\linewidth]{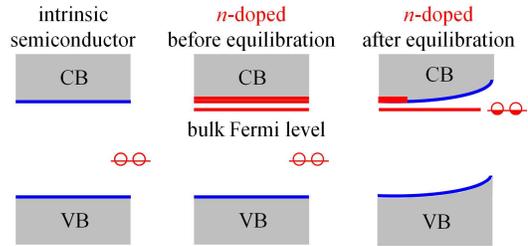}
\caption{\label{fig:schem} (color online) Schematic illustration of the electron transfer to acceptor states at a surface or interface of a \emph{n}-doped semiconductor (middle and right). In an undoped intrinsic semiconductor (left) no such electron transfer can take place resulting in an empty acceptor state in the band gap.}
\end{figure}

In HIOS research, first-principles approaches are indispensable due to the atomistic insight they provide. These calculations do typically not include the global effects of doping (i.e the position of the electron chemical potential or Fermi level that is controlled by doping). However, if donor or acceptor states are present at the interface of HIOS, the Fermi level position significantly affects the energy-level alignment  (cf Fig.~\ref{fig:schem}), as we will demonstrate with quantitative electronic-structure calculations in this Letter. A crucial aspect is the formation of a space-charge layer at surfaces and interfaces that gives rise to band bending. Since semiconductors are always intentionally or unintentionally doped, it is paramount to include doping explicitly in the theoretical description -- both the global effects as well as the formation of space-charge layers.

\begin{figure}[tbhp]
\includegraphics[width=0.8\linewidth]{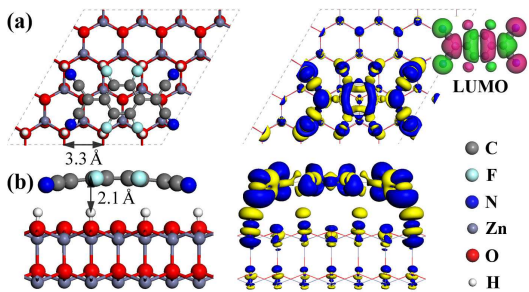}
\caption{\label{fig:F4_struct} (color online) Top (a) and side view (b) of F4TCNQ adsorbed on ZnO(000$\bar{1}$) (2$\times$1)-H (left) and the adsorption-induced electron density rearrangement for \emph{n}-doped ZnO (right). Electrons flow from the yellow to blue areas upon
adsorption. The electron accumulation region mimics the shape of the lowest unoccupied molecular orbital (LUMO) of the free F4TCNQ molecule.}
\end{figure}

To illustrate doping effects in HIOS, we consider the general problem of (organic) adsorbates on doped (inorganic) semiconductors, and investigate the properties of adsorbates as a function of the substrate doping concentration. We here adopt an approach for the calculation of defects in semiconductors~\cite{weinert_chalcogen_1986,scheffler_parameter-free_1988,Scherz1993}, which combines the statistical concept of a bulk Fermi level with atomistic first-principles calculations. In addition, we show how a space-charge layer, whose macroscopic dimensions far exceed the dimensions of supercells tractable in standard density-functional theory (DFT) calculations, can be properly accounted for. Then we apply the approach to DFT calculations of an example HIOS: a tetrafluoro-tetracyanoquinodimethane (F4TCNQ) monolayer on the ZnO(000$\bar{1}$) (2$\times$1)-H surface (see Fig.~\ref{fig:F4_struct}). We show that the doping in HIOS quantitatively affects interface properties such as the adsorption energy and electron transfer, or even qualitatively change the energy-level alignment at the interface. On \emph{n}-doped ZnO, F4TCNQ induces a large work function increase. This is accompanied by electron transfer that becomes vanishingly small in the limit of low bulk doping concentrations. Such a behavior has recently been demonstrated in photoemission measurements for F4TCNQ on ZnO~\cite{Schlesinger}.

A computational approach to describe doping effects for (organic) adsorbates on doped (inorganic) semiconductors should include (i) a Fermi level that depends on the bulk dopant concentration $N_\mathrm{D}$, (ii) electrons or holes that can be exchanged with the adsorbate, and (iii) the ensuing space-charge layer that leads to band bending. In the following  we demonstrate how to incorporate these three factors into a DFT-based framework.

Analogous to calculations of defects in the bulk or at interfaces~\cite{weinert_chalcogen_1986,scheffler_parameter-free_1988,Scherz1993}, excess electrons or holes are introduced into the semiconductor to model the global effects of doping. The adsorption energy ($\Delta E_q^{{\rm{ads}}}$) of an adsorbate that receives $q$ electrons from the electron reservoir with an electron chemical potential $\epsilon_{\rm{F}}$ can be written as \cite{SI}
\begin{align}
\Delta E_q^{{\rm{ads}}}({\epsilon_{\rm{F}}}) =  &(- E_q^{{\rm{surf/mol}}} + E_q^{{\rm{surf}}} + E_0^{{\rm{mol}}}) \nonumber \\ &+ (q\Delta {\epsilon_{\rm{F}}} - q\delta)  + {\Delta E^{{\rm{SC}}}}.
\label{eqn1}
\end{align}
$E_q^{{\rm{surf/mol}}}$ and $E_q^{{\rm{surf}}}$ are the total energies of the adsorbate system and the bare substrate computed in a supercell with $q$ excess electrons, and $E^{{\rm{mol}}}_0$ is the total energy of the neutral molecule. The second term in Eq.\ (\ref{eqn1}) quantifies the energy of the excess charge with respect to the electron reservoir: $\Delta {\epsilon_{\rm{F}}} = {\epsilon_{\rm{F}}} - {\epsilon_{\rm{CBm}}}$ when charging electrons ($q>0$) and $\Delta {\epsilon_{\rm{F}}} = {\epsilon_{\rm{F}}} - {\epsilon_{\rm{VBM}}}$ for holes ($q<0$), where $\epsilon_{{\rm{CBm}}}$ is the conduction band minimum (CBm) and $\epsilon_{{\rm{VBM}}}$ is the valence band maximum (VBM).
$q\delta$ is correction that accounts for the finite filling of the substrate's conduction (or valence) bands to an average energy $\delta$. $\delta$ depends on $q$ and reduces to zero in the limit of small $q$~\cite{SI}. In our calculations it never exceeds 0.2~eV. The last term in Eq.\ (\ref{eqn1}), $\Delta E^{{\rm{SC}}}$, denotes the energy correction for describing the space-charge layer.

Introducing excess charges into the unit cell is common practice in first-principles calculations of defects in the bulk~\cite{weinert_chalcogen_1986,scheffler_parameter-free_1988,Scherz1993,Walle-89,Zhang-91,van_de_walle_first-principles_2004,Persson-05}.  To keep the unit cell overall charge neutral and therefore to avoid a diverging Hartree energy, a uniform, compensating background of opposite charge is introduced. However, for surface calculations in the periodic slab approach such a homogeneous background resides also in the vacuum region and therefore builds up  a dipole with the original charge that is confined to the slab. This dipole and the associated energy diverge for increasing vacuum separations. To circumvent this problem we confine the compensating charge by applying the virtual-crystal approximation (VCA)~\cite{Vegard-21,Scheffler-87,Richter}. We modify the nuclear charge of semiconductor substrate atoms by a small amount $\Delta Z$~\cite{Scheffler-87,Richter,Moll:2013}, which results in corresponding excess electrons or holes in either the conduction or valence band. Independent tests show that in the limit of small $\Delta Z$ the VCA method provides a very reliable description of doping effects at surfaces~\cite{Richter}.

An important feature of semiconductor surfaces is that charge transfer from bulk dopants generates a space-charge layer and induces band bending. While a direct description of space-charge layers in first-principles calculations is computationally formidable because of the large length scales involved ($\sim$100 nm for ZnO with $N_{\rm{D}} = 10^{17}$ cm$^{-3}$), the effect can be taken into account using simple electrostatic considerations. For macroscopically extended  semiconductor surfaces, this electrostatic description is textbook knowledge~\cite{Sze-06}. The transfer of $q$ electrons (per surface supercell area $A$) from bulk dopants to the surface costs an energy of
\begin{equation}
E_1^{{\rm{SC}}}(q,{N_{\rm{D}}}) = \frac{{{e^2}}}{{6\varepsilon {\varepsilon _0}{N_{\rm{D}}}{A^2}}}{\left| q \right|^3},
\label{eqn2}
\end{equation}
where $e$ is the elementary charge, $\varepsilon$ the static dielectric constant and $\varepsilon_0$ the vacuum permittivity~\cite{Sze-06}.

Care has to be taken, however, because the DFT slab calculations also include a certain amount of band bending. The spatial extend of the space-charge layer is limited by the thickness of the slab ($d$) in the atomistic model. The concentration of excess charge in the supercell is then $N_{\rm{D}}^{\prime} = \left| q \right|/(Ad)$. To estimate the electrostatic energy of forming a space-charge layer within the slab, we apply Eq.\ (\ref{eqn2}) and replace $N_{\rm{D}}$ by $N_{\rm{D}}^{\prime}$:
\begin{equation}
 E_2^{{\rm{SC}}}(q) = \frac{{{e^2}d}}{{6\varepsilon {\varepsilon _0}A}}{q^2}.
\label{eqn3}
\end{equation}

We then take ${\Delta E^{{\rm{SC}}}} = -E_1^{{\rm{SC}}}(q,{N_{\rm{D}}}) + E_2^{{\rm{SC}}}(q)$ as the space-charge layer correction to the adsorption energy.
Equations (\ref{eqn2}) and  (\ref{eqn3}) demonstrate clearly that the electrostatic energy of a space-charge layer in a realistic semiconductor differs from that in the DFT slab calculations by its $q$-dependence. However, with the exception of Ref.~\onlinecite{Richter} no such correction term has been taken into account in electronic-structure studies so far. As we will demonstrate, the correction affects the predicted electron transfer and adsorption energy considerably, and thus is essential in DFT studies of HIOS. Since the inclusion of ${\Delta E^{{\rm{SC}}}}$ permits us to decouple $N_{\rm{D}}$ from the excess electrons or holes introduced in the supercell, we rewrite the adsorption energy as a function of $N_{\rm{D}}$:
\begin{align}
&\Delta E_q^{{\rm{ads}}}({N_{\rm{D}}}) =   (- E_q^{{\rm{surf/mol}}} + E_q^{{\rm{surf}}} + E_0^{{\rm{mol}}}) \nonumber \\
& + (q\Delta {\epsilon_{\rm{F}}}({N_{\rm{D}}}) - q\delta)
- \frac{{{e^2}}}{{6\varepsilon {\varepsilon _0}{N_{\rm{D}}}{A^2}}}{\left| q \right|^3} + \frac{{{e^2}d}}{{6\varepsilon {\varepsilon _0}A}}{q^2}.
 \label{eqn4}
\end{align}
The $N_{\rm{D}}$ dependence of the Fermi level [$\Delta \epsilon_{\rm{F}}(N_\mathrm{D})$] is known for many semiconductors and is described in the Supplemental Material~\cite{SI} for ZnO.

Next, we apply our approach to a F4TCNQ monolayer on the ZnO(000$\bar{1}$) surface (shown in Fig.~\ref{fig:F4_struct}). F4TCNQ is a strong electron acceptor, that is commonly used for surface/interface modifications and work function tuning~\cite{Koch-05,Romaner-07,Rangger-09,Chen-09,Schlesinger}. ZnO is a suitable inorganic component in HIOS, and is natively \emph{n}-doped, presumably due to defects like oxygen vacancies, zinc interstitials or hydrogen~\cite{Ozgur-05,van2000hydrogen,janotti2006hydrogen}. The oxygen-terminated ZnO(000$\bar{1}$) surface has been extensively studied~\cite{Woell-07,Meyer-04,Lauritsen-11}. The ZnO(000$\bar{1}$) (2$\times$1)-H phase, in which every second row of surface O atoms is decorated with H atoms, is the thermodynamically most stable structure at typical experimental growth conditions, according to our and previous studies~\cite{Meyer-04,Lauritsen-11,Moll:2013}. We also showed previously that ZnO(000$\bar{1}$) surfaces with lower hydrogen concentrations (less than 50\%) may be stabilized by \emph{n}-type bulk dopants in hydrogen-deficient environments~\cite{Moll:2013}. However, this does not affect our conclusions. For simplicity, we assume hydrogen-rich environments and adopt the ZnO(000$\bar{1}$) (2$\times$1)-H surface as model for the bare surface prior to F4TCNQ adsorption.

The DFT calculations were performed using the all-electron full-potential code FHI-aims~\cite{Blum-09}. We applied the Heyd-Scuseria-Ernzerhof (HSE) hybrid functional~\cite{Heyd/Scuseria/Ernzerhof:2003,Krukau-06}, but  the admixture of exact-exchange was adjusted to 50\% (denoted HSE*) as in Ref.~\onlinecite{Ramprasad-12}, instead of the default value of 25\% to achieve the best compromise between the experimental bandwidth, the band gap and the energetic ordering for the two subsystems. With 4.3~eV the band gap of ZnO is then overestimated compared to the experimental value of 3.44~eV at zero temperature \cite{Cardona/Thewalt:2005}. Equilibrium geometries were calculated with the Perdew-Burke-Ernzerhof (PBE) functional~\cite{Perdew-96} combined with screened van der Waals (vdW) corrections~\cite{Zhang-11} (PBE+vdW$^{\mathrm{scr}}$) excluding vdW interactions within the ZnO substrate. Tests for
HSE*+vdW$^{\mathrm{scr}}$ show negligible geometry differences. The electronic structure and adsorption energies were calculated using HSE*+vdW$^{\mathrm{scr}}$ \cite{note1}.

On the ZnO(000$\bar{1}$) (2$\times$1)-H surface, the four CN groups of F4TCNQ interact attractively with the surface H atoms and repulsively with the surface O atoms. This results in a stable geometry (see Fig.~\ref{fig:F4_struct}), in which F4TCNQ lies face-on on the substrate with the cyano groups located above the surface H atoms. The molecule distorts slightly upon adsorption, placing the N atoms 0.4 \AA \  below the F atoms. We also found some metastable geometries at other adsorption sites, whose energies are at least 0.5 eV higher. The monolayer morphology depends on the molecular coverage. At low coverages, the F4TCNQ molecules tend to be well separated from each other due to intermolecular repulsions, even in the limit of zero electron transfer. At high coverages, the intermolecular repulsion has a more direct influence on the packing motif. Here we focus on the coverage of 0.67 molecule/nm$^2$, which is the tightest monolayer packing that is commensurate with the H-overlayer and was used to study F4TCNQ on coinage metals~\cite{Romaner-07,Rangger-09}. Figure~\ref{fig:F4_struct} shows the most stable monolayer structure we found~\cite{note2}.

\begin{figure}[tbhp]
\includegraphics[width=0.85\linewidth]{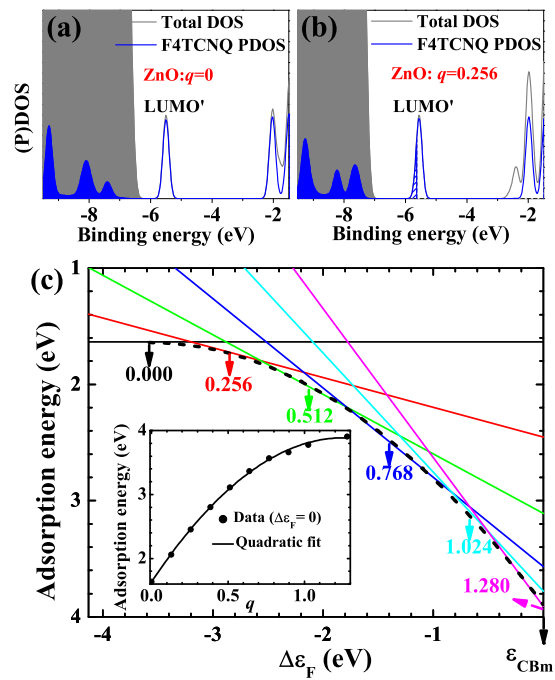}
\caption{\label{fig:adeng} (color online) (a,b) Calculated total density of states (DOS) and projected DOS (PDOS) onto F4TCNQ for intrinsic ($q = 0$)  and electron-doped ($q = 0.256$) ZnO, where $\Delta Z = q/128$. The position at which the LUMO$'$ pins in (b) corresponds to the Fermi level at the surface, which is determined by the space-charge layer. The binding energy is referenced to the vacuum level. (c) Adsorption energy as a function of $\Delta {\epsilon_{\rm{F}}} = {\epsilon_{\rm{F}}} - {\epsilon_{{\rm{CBm}}}}$ for different charge states $q$ (indicated by the corresponding numbers), obtained by Eq.\ (\ref{eqn1}) excluding ${\Delta E^{{\rm{SC}}}}$. The inset shows data points of the adsorption energy versus $q$ at $\Delta {\epsilon_{\rm{F}}} = 0$ and a quadratic fit. The fit then gives the adsorption energy as a function of $\Delta {\epsilon_{\rm{F}}}$ (dashed line in the main figure), which also exhibits a quadratic dependence.}
\end{figure}

Figure~\ref{fig:adeng} summarizes our results based on Eq.~\ref{eqn1}, but without the space-charge correction. For undoped calculations the LUMO$'$ (LUMO after adsorption) of F4TCNQ lies in the band gap and is unoccupied [see Fig. \ref{fig:adeng}(a)]. 
As soon as excess electrons are offered (i.e. $q>$0), these are immediately transferred to the LUMO$'$ of F4TCNQ [see Fig. \ref{fig:adeng}(b)], which is further evidenced by the adsorption induced charge rearrangement (see Fig.~\ref{fig:F4_struct}). As a result the work function increases. Due to the linear term $q\Delta \epsilon_{\rm{F}}$ in Eq.~\ref{eqn1}, calculations for different $q$ manifest themselves in lines with different slopes in Fig.~\ref{fig:adeng}. For a given Fermi energy, the line with the lowest energy indicates how much charge is transferred to F4TCNQ. Figure~\ref{fig:adeng} illustrates that the adsorption energy depends quadratically on the Fermi energy. Such a quadratic behavior is expected from a simplified parallel capacitor model for the charge transfer between the substrate and the adsorbate. The DFT results therefore show that i) the electron transfer and adsorption energy increase with increasing Fermi level and that ii) undoped calculations (i.e. the majority of all surface calculations in the literature) do not capture this effect and predict zero electron transfer and vacuum level alignment (cf Fig.~\ref{fig:schem}).

If the Fermi level position at the surface is known experimentally, the amount of electron transfer and the corresponding adsorption energy can be read off Fig.~\ref{fig:adeng}, once the data has been corrected for the erroneous space-charge layer present in the slab calculations. To proceed, we include the space-charge layer correction using Eq.~\ref{eqn4}. For a given $N_{\rm{D}}$, we maximize $\Delta E_q^{{\rm{ads}}}({N_{\rm{D}}})$ with respect to $q$, which then gives the optimal electron transfer per molecule $Q$ and the associated adsorption energy $\Delta E^{{\rm{ads}}}$.

\begin{figure}[tbhp]
\includegraphics[width=0.85\linewidth]{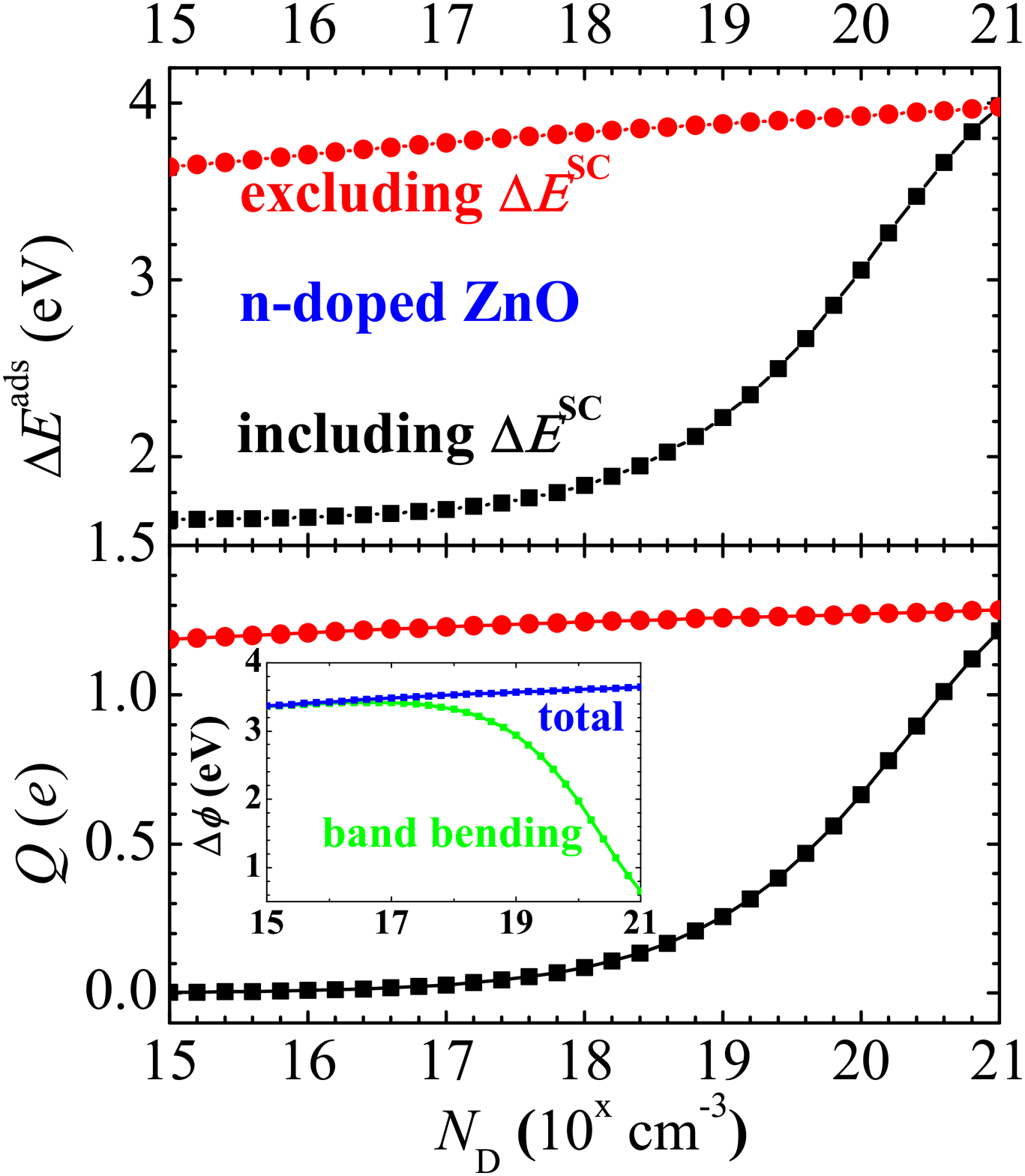}
\caption{\label{fig:Q/Eng_vs_ND} (color online) Adsorption energy $\Delta E^{{\rm{ads}}}$ and electron transfer $Q$ as a function of (\emph{n}-type) dopant concentration $N_\mathrm{D}$ for F4TCNQ/ZnO(000$\bar{1}$) (2$\times$1)-H, calculated excluding and including the space-charge layer correction ${\Delta E^{{\rm{SC}}}}$. The F4TCNQ-induced work function change $\Delta \phi$ and the associated band-bending contribution (from calculations including ${\Delta E^{{\rm{SC}}}}$) are shown in the inset.}
\end{figure}

The result is shown in Fig.~\ref{fig:Q/Eng_vs_ND}, which  summarizes the main message of this Letter. Both the electron transfer and adsorption energy exhibit a pronounced dependence on $N_{\rm{D}}$. It is well known that the magnitude of band bending is inversely proportional to $N_\mathrm{D}$. Therefore, for low $N_\mathrm{D}$ band bending alone can lift up the LUMO$'$ to the Fermi energy, inducing a large work function increase (inset of Fig.~\ref{fig:Q/Eng_vs_ND}). This reduces the required electron transfer to nearly zero and the adsorption energy assumes the value of 1.6~eV we find in the undoped calculation. As $N_{\rm{D}}$ increases, the work functions before and after adsorption only slightly vary, and the work function change depends weakly on $N_\mathrm{D}$ (inset of Fig.~\ref{fig:Q/Eng_vs_ND}). While band bending reduces, the electron transfer picks up. In the process, the adsorption energy more than doubles. For heavily \emph{n}-doped ZnO used in transparent conductors, the adsorption energy has increased by more than 2~eV to a value of 4.0~eV. For comparison, without the space-charge layer correction the DFT results (red line in Fig.~\ref{fig:Q/Eng_vs_ND}) miss the $N_\mathrm{D}$ dependence entirely and only give reasonable results in the high-doping region.

Real HIOS interfaces are typically not as ``ideal'' as the ones discussed here. Realistic models would have to include not only the spatial profile of the dopants, but also information on other impurities at or near the interface (e.g. oxygen vacancies~\cite{Ozgur-05}) that could pin the Fermi level at defect levels~\cite{mosbacker2005role} and limit the amount of band bending. In some cases the ZnO films may be thinner than the space-charge layer or ZnO nanoclusters or nanocolumns are used. Then we expect doping effects to be film-thickness/structure-size dependent. All these issues could be important for HIOS, but their resolution requires input from experiments.

Finally, we make contact with recent photoemission experiments for F4TCNQ on ZnO(000$\bar{1}$)~\cite{Schlesinger}. We predict for \emph{n}-doped ZnO that the work function increases up to around 5.7 eV upon adsorption, due to the partial occupation and pinning of the LUMO$'$ at the Fermi level. This final work function is insensitive to variations in the surface termination (i.e. the hydrogen deficiency alluded to before), because it is determined by the distance of the LUMO$'$ to the vacuum level above the F4TCNQ film and thus independent from the position of the LUMO$'$ in the band gap. At the experimental $N_\mathrm{D}$ of approx. 10$^{17}$ cm$^{-3}$, Fig.~\ref{fig:Q/Eng_vs_ND} predicts a vanishing electron transfer of 0.03 electrons/molecule (or 0.02 electrons/nm$^{2}$). The results are consistent with the photoemission measurements, which observe a final work function of 5.9 eV and no noticeable electron transfer~\cite{Schlesinger}. While band bending dominates the work function change in our theoretical description, band bending is limited to 0.5~eV in experiment~\cite{Schlesinger}, possibly because of a different interface structure and pinning at deep defect states.

It is in principle possible to change the level alignment at a HIOS interface from Fermi-level pinning to vacuum level alignment. For the example of F4TCNQ on ZnO(000$\bar{1}$) (2$\times$1)-H, we demonstrate that such a transition occurs when the substrate doping varies from \emph{n} to \emph{p}-type. This behavior is expected to be quite general in HIOS and it would be interesting to experimentally test also materials like GaN or Si. Moreover, we show that the amount of electron transfer and therefore the amount of trapped charge at the interface significantly depend on the bulk doping concentration. The trapped charges can act as scattering centers and affect transport properties at the interface. Therefore, high bulk doping concentrations for improved charge injection/transport have to be balanced against the resulting interface charges for device optimization.

\begin{acknowledgements}
This work is supported by the DFG collaborative research project 951 ``HIOS''. Y. Xu acknowledges support from the Alexander von Humboldt foundation.
\end{acknowledgements}

%

\end{document}